\documentclass[preprint2]{aastex}

\def\atwo{\mbox{$A_{wb}$}}
\def\dcs{\mbox{$D_{cs}$}}
\def\dcb{\mbox{$D_{cb}$}}
\def\dhs{\mbox{$D_{hs}$}}
\def\eD{$\epsilon_D$}
\def\ee{\mbox{e}}
\def\eg{e.g.,}
\def\gyr{\mbox{Gyr}}
\def\kcb{\mbox{$K_{cb}$}}
\def\kwb{\mbox{$K_{wb}$}}
\def\khb{\mbox{$K_{hb}$}}
\def\kms{\mbox{km/s}}
\def\kpc{\mbox{kpc}}

\def\msun{\mbox{M$_\odot$}}
\def\Msun{\mbox{M$_\odot$}}

\def\nbody{\mbox{$N-$body}}

\def\omp{\mbox{$\Omega_p$}}

\def\rv{\mbox{$R_{\rm vir}$}}

\def\sech{\mbox{sech}}

\def\mathnew{\mathsurround=0pt}
\def\simov#1#2{\lower .5pt\vbox{\baselineskip0pt
    \lineskip-.5pt\ialign{$\mathnew#1\hfil##\hfil$\crcr#2\crcr\sim\crcr}}}  

\def\simless{\mathrel{\mathpalette\simov <}}
\def\'#1{\ifx#1i{\accent"13\i}\else{\accent"13#1}\fi}

\begin{document}

\title{Bars and Cold Dark Matter Halos}

\author{Pedro Col\'in}
\affil{Centro de Radiostronom\'ia y Astrof\'isica, Universidad Nacional Aut\'onoma de M\'exico, 
Apartado Postal 72-3 (Xangari), 58089 Morelia, Michoac\'an, Mexico}

\author{O. Valenzuela}
\affil{Department of Astronomy, Box 351580, University of Washington, Seattle, WA 98195-1580}

\and 

\author{A. Klypin}
\affil{Department of Astronomy, Box 30001, Department 4500, New Mexico State University, 
Las Cruces, NM 88003-0001}


\begin{abstract}
The central part of a dark matter halo reacts to the presence and
evolution of a bar. Not only does the halo absorb angular momentum from
the disk, it can also be compressed and have its shape modified. We
study these issues in a series of cosmologically motivated, highly
resolved N-body simulations of barred galaxies run under different
initial conditions. In all models we find that the inner halo's central
density increases. We model this density increase using the standard
adiabatic approximation and the modified formula by Gnedin et al.
and find that halo mass profiles are better reproduced 
by this latter. In models with a strong bar, the dark
matter in the central region forms a bar-like structure (``dark matter
bar''), which rotates together with the normal bar formed by the
stellar component (``stellar bar''). The minor-to-major axial ratio of a
halo bar changes with radius with a typical value 0.7 in the central
disk region. DM bar amplitude is mostly a function of the stellar bar
strength. Models in which the bar amplitude increases or stays
roughly constant with time, initially large (40\%-60\%) 
misalignment between the halo and disk
bars quickly decreases with time as the bar grows. The halo bar is
nearly aligned with the stellar bar ($\sim 10^\circ$ lag for the halo) after
$\sim 2\ \gyr$.  The torque, which the halo bar exerts on the stellar bar,
can serve as a mechanism to regulate the angular momentum transfer
from the disk to the halo. 
\end{abstract}

\section{Introduction}

The cold dark matter scenario for galaxy formation predicts 
that galaxies should be embedded within massive, extended, hot, 
and cuspy dark matter halos. The halos can have as much 
as twenty times more mass than the disk. They extend well beyond the
visible galaxy. Halos are hot because they are supported against gravity by
a large velocity dispersion. Halos are  cuspy because their central density
profiles go as $r^{-\alpha}$, with $\alpha \sim 1.0$. Unfortunately,
for historical reasons\footnote{A well established {\it galactic-scale} 
cold dark matter cosmogony did not arise but until late 1990s.},
for simplicity, and because of lack of computer resources,  
most studies of stellar bars have assumed unrealistic halos; from 
rigid ones (papers in the 1970s, with few exceptions) 
to live but small ones (papers in the 1980s and 1990s). The need for
a realistic halo component is being recognized and nowadays models in
which this dark component satisfies some or all of the above requirements 
are more common to find in the literature
\citep{ds00, lia03, vk03, od03, hbk05}.

A {\it systematic} study of the structural and dynamical changes that
the dark matter (DM) component experiences in the presence and 
during the evolution of a bar is missing.  Most studies that address
the issue have focused on the angular momentum transfer from the bar
to the halo \citep[\eg][]{weinberg85, combes90, lc91, hw92, lia96,
ds00}.  For example, \citet{weinberg85} and \citet{hw92}, using rigid
bars, found that a bar loses its angular momentum in few rotation
periods because of the dynamical friction that the dark matter exerts
on it. This angular momentum transfer flattens out the initially cuspy
halo density profile \citep[see also][]{hbk05}.  

There is no doubt that the disk loses angular momentum due to this 
transfer mechanism, but the
amount and the rate at which this happens is still a matter of
debate. Fully self-consistent N-body simulations with live disk and
dark matter halos \citep[\eg][]{lia96, ds98,ds00, am02, vk03} show
that bars slow down far less than predicted by \citet{weinberg85} and
\citet{hw92}. \citet{ds98,ds00} find in their massive halo models,
i.e., those for which the contributions of the disk and the halo to the
mass in the central region are comparable, that the disk loses about
40\% of its initial angular momentum in $\sim 10\ \gyr$. During this
time interval the bar pattern speed, \omp, decreases by a factor of
four. \citet{vk03}, in simulations with much better force resolution
and a more realistic cosmological halo setup, find a decrease in \omp\
of only a factor of two in $\sim 7\ \gyr$ (see their model $A_2$).

The response of the DM halo to the transfer of the stellar material
from the middle parts of the disk closer to the center has been
studied recently by \citet{hbk05} and \citet{sell03} in relation to
the {\it cuspy} problem. Cosmology predicts cuspy inner density
profiles that appear to disagree with those derived from rotation
curves of dwarf and low surface brightness galaxies
\citep[\eg][]{deBlok, WdBW2003}, but see also
\citet{Swaters2003,rhee04}. \citet{hbk05} find a decrease in the halo
central density, while simulations by \citet{sell03} show the
contrary. Here we discuss halo density profiles within the context of
adiabatic approximation models
\citep[\eg][]{blumenthal86,rg87,gkkn04}. Do dark matter halos respond
to the increase of stellar mass in the center by contracting? If so,
can this increase in density be modeled by the standard adiabatic
approximation? In agreement with \citet{sell03}, we will show that in
our models the halo central density increases, sometimes by as much as
a factor of 3.

The influence of the bar on the morphology of a halo or a spheroid is
a point that has only been touched briefly in the literature. Yet, it
is an issue that deserves a more detailed analysis. It may be
important for accurate testing of cosmological predictions, which are
becoming a reality \citep[\eg][]{Ibata01}.  For example, in our bar
models we find that the initially spherical halo flattens to values
comparable to those found in cosmological halos, for which the
minor-to-major axial ratio $s$ is $s \approx 0.7$
\citep[\eg][]{js02}. This correlation between the morphology of the
stellar component and that of the DM halo has been also
seen in cosmological gasdynamic simulations of galaxies
\citep{stelios04}.

Returning to bar simulations, \citet{combes90} found no significant
flattening of the bulge particles ($s \approx 1$) in their experiment A,
for which $M_b/M_d = 0.5$, where $M_b$ and $M_d$ are the mass of the
bulge and the disk, respectively.  \citet{ds00} studied the halo
response to the bar by measuring the second harmonic of the angular DM
particle distribution and found that the phase difference between the
disk and the halo bar gradually goes to zero. This effect was also
seen in the simulations by \citet{od03}: after $t \sim 10\ \gyr$ the
orientations of bars (halo and disk) stays within 10$^\circ$ of one
another. The halo bar found by \citet{od03} was not very
elongated. The axial ratio was only 0.88. Halo bars have also seen by
\citet[][ see their Fig. 2]{hbk05} and \citet{lia05a,lia05b}.  Based
on preliminary results of simulations with strong bars,
\citet{lia05a,lia05b} finds that halo bars are prolate-like, with
inner axial ratios being 0.7 or 0.8. Contrary to results of other
groups, her halo bars are roughly aligned to the disk ones at all
times.

More recently, in order to understand the role of the halo in the
formation and evolution of the bar, \citet{lia02,lia03} has carried
out an orbital-resonant study of the disk and halo particles in models
in which the disk parameters are kept fixed while varying the degree
of the halo influence.  She found significant differences in the halo
orbital properties between her halo and disk-dominated models.  The
inner Linblad resonance, ILR, is present in the halo-dominated models,
while in the disk-dominated ones is not. In general, she found that
there is much more material between the ILR and corotation in the
former models than in the latter ones. Because these halo-dominated
models have stronger bars than their counterparts- disk-dominated ones,
this result shows the positive effect that a live halo has on the bar
growth.

The paper is organized as follows:  In section 2 we introduce the
models to be explored in this paper. In Section 3 we present the
evolution of the amplitude and pattern speed of the bar. Section 4 is
devoted to an analysis of models of adiabatic compression, which are used in the
context of the inner disk density growth. The influence of the disk
and stellar bar on the initially spherically symmetric halo is discussed
in section 5. In particular, we will show that once formed, the halo bar 
tends to rapidly align with the stellar bar. A brief discussion
of the redistribution and evolution of the disk angular momentum is
given in section 6. In section 7 we discuss our results and summarize
our main conclusions.

\begin{planotable}{lccccccc}
\tablecolumns{8}
\tablewidth{0pc}
\tablecaption{Simulation Parameters}
\tablehead{\colhead{Parameter} & \colhead{\atwo} & \colhead{\dcs} & \colhead{\dcb} & 
\colhead{\dhs} & \colhead{\kcb} & \colhead{\kwb} & \colhead{\khb} }
\startdata
 Disk mass (10$^{10}$ \msun)       & 4.28 & 5.0   & 5.0   & 5.0   & 5.0   & 5.0   & 5.0   \\
 Total mass (10$^{12}$ \msun)      & 2.04 & 1.43  & 1.43  & 1.43  & 1.43  & 1.43  & 1.43  \\
 Disk exponential length $R_d$(kpc)& 3.50 & 2.57  & 3.86  & 2.57  & 3.86  & 3.86  & 3.86  \\
 Disk exponential height $z_0$(kpc)& 0.14 & 0.20  & 0.20  & 0.20  & 0.20  & 0.20  & 0.20  \\
 Toomre parameter $Q$              & 1.6  & 1.3   & 1.3   & 1.8   & 1.2   & 1.5   & 1.8   \\
 Halo concentration $C$            & 15   & 17    & 17    & 17    & 10    & 10    & 10    \\
 Number of disk particles (10$^5$) & 2.0  & 2.33  & 2.33  & 4.60  & 2.33  & 2.33  & 2.33  \\
 Total Number of particles (10$^6$)& 2.7  & 2.5   & 1.9   & 3.80  & 2.3   & 2.7   & 2.3  \\
 Particle mass (10$^5$ \msun)      & 2.14 & 1.07  & 2.14  & 1.07  & 2.14  & 2.14  & 2.14  \\
 Time-step  (10$^4$ yr)            & 1.5  & 2.0   & 1.5   & 1.5   & 1.2   & 1.5   & 2.0   \\
 Minimum cell size (pc)            & 22   & 48    & 22    & 22    & 24    & 22    & 48   \\
 Disk-to-DM mass ratio inside $R_d$& 0.82 & 2.03  & 0.98  & 2.03  & 1.95  & 1.95  & 1.95  \\
 Strong Bar after 5~\gyr ($A_2>0.4$)& yes  & yes   & yes   & no    & no    &  no   & no   \\       
\enddata
\end{planotable}

\section{Models and simulations}

\subsection{Initial conditions}
\begin{figure}[htb!]
\plotone{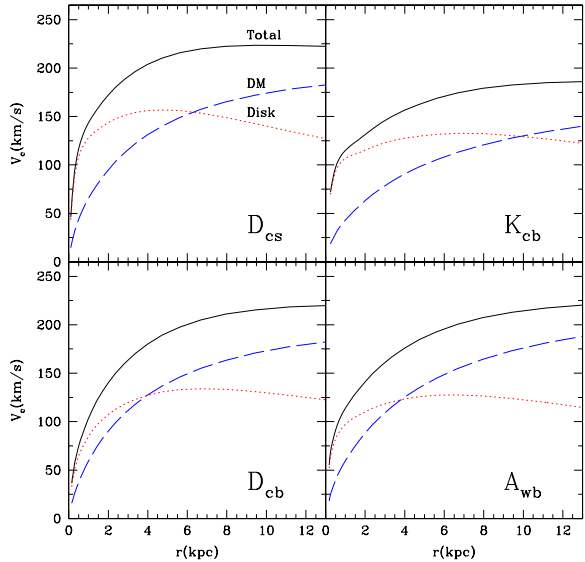}
\caption{Initial circular velocities profiles for models
\dcs\ and \kcb\ (top panels), and \dcb\ and \atwo\ (bottom
panels). Total, stellar, and halo components are denoted 
by solid, dotted, and dashed lines, respectively. In the
disk-dominated models (top panels) the system evolves secularly 
producing a weak bar at later stages, while in the halo-dominated models 
(bottom panels) a strong bar is present after 5~\gyr\ of evolution.} 
\end{figure}

The initial conditions are described in \citet{vk03}. Here we briefly
summarize parameters of our models. The system of a halo and a disk, with no initial bulge
or a bar, is generated using the method of \cite{hern93}. In
cylindrical coordinates the density of the stellar disk is
approximated by the following expresion:
\begin{equation}
\rho_d (R,z) = \frac{M_d}{4\pi z_0 R_d^2} \ee^{-R/R_d} \sech^2(z/z_0),
\end{equation}
where $M_d$ is the mass of the disk, $R_d$ is the scale length, and
$z_0$ is the scale height. The latter is assumed constant through
the disk. The radial and vertical velocity dispersion are given by
\begin{equation}
\sigma(R)= Q \frac{3.36 G \Sigma(R)}{\kappa(R)}, \hspace{0.5cm}
\sigma_z^2 (R) = \pi G z_0 \Sigma(R),
\end{equation}
where $\kappa$ is the epicyclic frequency, and $Q$ the stability or
Toomre parameter. Our models keep $Q$ fixed along the disk. 
The azimuthal velocity and its dispersion are found using the asymmetric
drift and the epicycle approximations. 

The models assume a NFW density 
profile \citep{nfw97} for the halo component,
which is described by 
\begin{eqnarray}
\rho_{DM}(r) = \frac{\rho_0}{x(1+x)^2}, \hspace{0.5cm} x \equiv r/r_s, \\
M_{vir} = 4\pi \rho_0 r_s^3 \left[ \ln(1+c) - \frac{c}{1+c} \right],  \hspace{0.2cm} c =  
\frac{R_{vir}}{r_s},
\end{eqnarray}
where $M_{vir}$, $R_{vir}$, and $c$ are the virial mass, the virial radius, 
and concentration of the halo, respectively. Given $M_{vir}$, the virial 
radius is found once a cosmology is adopted\footnote{We adopt the 
flat cosmological model with a non-vanishing cosmological constant 
with $\Omega_0=0.3$ and $h=0.7$.}. The equation (4-56) of \citet{BT}
and the assumption of isotropy in the velocities allow us 
to determine the radial velocity dispersion as
\begin{equation}
\sigma^2_{r,DM} = \frac{1}{\rho_{DM}} \int_r^\infty \rho_{DM} \frac{GM(r)}{r^2} dr,
\end{equation}
where $M(r)$ is the mass contained within radius $r$ and $G$
the gravitational constant. 
\begin{figure*}[p]
\includegraphics[width=\textwidth]{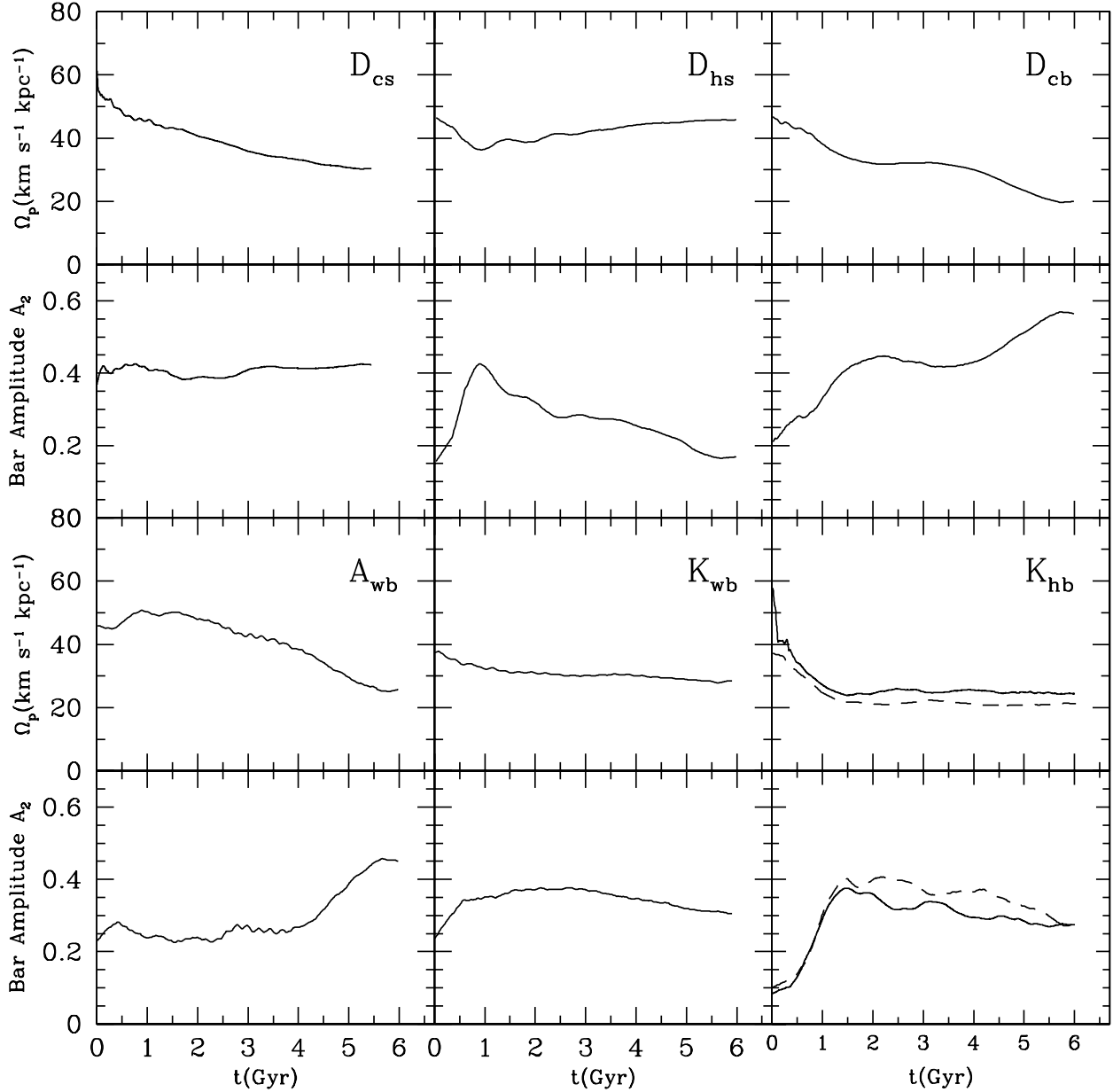}
\caption{Evolution of the bar pattern speed (first and third rows) and 
the bar amplitude (second and fourth rows) for six of our seven models.
Models \dcs, \kwb, and \khb\ 
evolve very slowly after the bar formation. In models
\dhs\ and \kcb\ (not shown in the plot), the bar amplitude, after reaching a peak, 
decreases as a function of time. In models \atwo\ and \dcb,
at late stages of the evolution, the bar develops a second growth period.
The dashed curves in the \khb\ model show results obtained with the GADGET code.}
\end{figure*}

The dark matter halo is truncated at the virial radius, which for our
$D$ and $K$ models is 287 kpc, while for our \atwo\ model it is 323 kpc
(see Table 1). The disk component is realized with particles of
equal mass. The halo is composed of particles of different mass placed
inside out as we go from less to more massive particles. The smallest
particles have a mass equal to those of the disk. The multiple-mass
scheme is designed to reduce the total number of particles and, thus,
the CPU time.

Parameters of our models are presented in Table~1. They 
were chosen to cover a relatively wide range of the 
parameter space: there are hot and cold
models, \khb\ versus \kcb;  there is a disk-dominated \dcs\ 
and a sub-maximum \dcb\ model.
There are models with different DM halo concentrations: \dcb\ and \kcb. 
The parameter values are similar to those  used by \citet{vk03}, 
whose models were chosen to mimic the properties of the Milky Way galaxy \citep{kzs02}.
More specifically, models labeled with $D$ have all the same disk mass, total mass, and 
halo concentration. This latter is above the median $\sim 13$ for a halo of this
mass and cosmology adopted here \citep{Bullock2001}. On the other hand, models
$K$ differ one from other only by the value of the disk stability parameter $Q$.

As a measure of how dense is the
halo in the central region we show in the Table 1 the disk-to-DM mass
ratio evaluated at the disk exponential length $R_d$. The values are
about 1--2. In other words, from the beginning the dark matter is
subdominant in the central region. This is broadly consistent with the
mass modeling of the Milky Way galaxy \citep{kzs02}: even for
cosmologically motivated cuspy DM halos, the density of the dark
matter must be smaller than the density of the baryons in the central
region of galaxies.  This is different from what is often assumed in
simulations of barred galaxies. For example, in DM dominated
models of \citet{am02} the halo mass dominates everywhere inside
$2R_d$. This is not consistent with cosmological predictions.
We do allow variations of the DM mass in the central region.
For example, the model \dcb\ has a denser halo as compared with models $K$.
Models \dcb\ and \atwo, in which initially the disk is less dominant, 
develop strong bars.  The last row of Table~1 classified the models according to stellar bar
strength after 5~\gyr\ of evolution. We label them as strong bar
models if they develop a bar with an amplitude $A_2$ larger than
0.4. 

Figure~1 shows the initial circular velocities profiles $V_c =
\sqrt{GM(r)/r}$ for all the components.  This estimate of $V_c$
assumes spherical symmetry. The circular velocity estimated in the
plane of the disk using the real non-spherical distribution $V_{{\rm
c,}z=0}\equiv \sqrt{g(r) r}$, where $g(r)$ is the gravitational
acceleration in the disk plane, is very close (whithin 5 percent) to
the circular velocity $V_c$ at radii larger than 1\ \kpc. At smaller
distances it is slightly (10-20\%) {\it below} $V_c$.

\subsection{Numerical simulations}

The simulations were run with the Adaptive Refinement Tree (ART) code 
\citep{KKK97}. 
 The ART code
starts with a uniform grid, which covers the whole computational
box. This grid defines the lowest (zeroth) level of resolution of the
simulation. 
Two grids with $256^3$ and $128^3$ cells in the zeroth-level 
and boxes of 1.43~Mpc and 1.0~Mpc across, respectively,
were used by the simulations presented here.  The models are placed at the
center of the grid and far from the periodic images. Because the size
of the models are small as compared with the box sizes, the
effects of the periodical images are negligible. For example, at a distance 
of 100~kpc from the center of a model, in the 1.43 Mpc box,
the relative contribution of periodic images is less that
$7 \times 10^{-7}$ of the main galaxy force. Even at 250~kpc, which is
close to the virial radius for the models, the tidal force is only
$6 \times 10^{-5}$ of the force from the central image.  Bar formation
and evolution are practically unaffected by the periodic boundary
conditions.

 The ART code achieves high spatial resolution by 
refining the base uniform grid in all high-density regions with an automated 
refinement algorithm.
 The standard Cloud-In-Cell algorithm is used to compute
the density and gravitational potential on the zeroth-level mesh with
periodical boundary conditions.  The code then reaches high force
resolution by refining all high density regions using an automated
refinement algorithm.  The refinements are recursive. A refined region
can also be refined. Each subsequent refinement level has half of the
previous level's cell size. This creates a hierarchy of refinement
meshes of different resolution, size, and geometry covering regions of
interest. Because each individual cubic cell can be refined, the shape
of the refinement mesh can be arbitrary and effectively match the
geometry of the region of interest. This algorithm is well suited for
simulations of a selected region within a large computational box, as
in the simulations presented below.

The criterion for refinement is the local density of particles. If the
number-density of particles in a mesh cell (as estimated by the Cloud-In-Cell
method) exceeds the level $n_{\rm thresh}$, the cell is split
(``refined'') into 8 cells of the next refinement level.  The
refinement threshold depends on the refinement level. The threshold
for cell refinement was low on the zeroth level: $n_{\rm
thresh}(0)=2$.  Thus, every zeroth-level cell containing two or more
particles was refined.  The threshold was higher on deeper levels of
refinement: $n_{\rm thresh}=3$ and $n_{\rm thresh}=4$ for the first
level and higher levels, respectively. Note that the code actually
does not count the number of particles in a cell to decide whether the
cell should be split or not. It uses the density field, which is less
noisy than counting particles. It then filters the density field
to make  the map of refinement even smoother. On average the algorithm
maintains 1-4 particles per cell.  Yet, some cells have few particles
in them and some can be even empty. Those cells are not used for
estimates of the gravitational acceleration acting on particles.  For
each particle the code checks whether all 56 cells surrounding the cell
hosting the particle are on the same level of refinement. If they are
not, then the code assigns the particle to a lower resolution and
checks again for the refinement level of all its neighbors.
In two of our models (\dcs\ and \khb) we limited the resolution
 to not more than 7 levels of refinement.

The resolution changes from place to place and it also changes with
time.  During the initial stage of evolution -- the formation of the
bar -- the density in the central part increases substantially. For
example, in the model \kcb\ the density at 250~pc increases by a factor
of 10 by the end of the evolution. The code reacts to this density
increase by changing the number of high-resolution cells. The lion's
share of this increase happens during the first 0.5~Gyr when the
system develops very strong non-axisymmetric fluctuations. For example, in the case
of the model \kcb\ the density at the central 250~pc increases 6~times
during first $t=0.5$~Gyr. It will take another 5~Gyr to double it. The
code increases the number of high resolution 24~pc cells 14 times
during the first 0.5~Gyr of evolution. The number of cells 
then changes only by 30
percent during the rest 5~Gyrs. The increase on lower 48~pc level of
resolution is even smaller: the number of cells increased only by 20
percent during the whole evolution. Note that this implies that the size of the
high resolution region  changes only by 6 percent.  In other
words, the increase of the resolution with time is very mild and most
of it happens before the bar forms and settles. 

The spatial resolution varies with the distance. Typically in the
models the central roughly elliptical region with large axis 1.5~kpc
(4~kpc) is resolved with the 20-25~pc (40-50~pc).  The whole disk (up
to radius of about 15~kpc) is resolved with 80-100~pc cells.  The
force resolution in our simulations is twice larger than a cell
size. Inside one resolution element (roughly a sphere of two cell
radii) the code has on average 64 particles. This large number of
particles implies that the code does not produce too high resolution
for the number of particles present in our models.  The effect of
close encounters is actually even smaller than what it seems from
naive counting particles. The natural scale of the close encounters is
Chandrasekhar's minimum impact parameter $b_{\rm min}= Gm/V^2$, which
is the distance for 90 degree deflection (here $m$ is the particle
mass and $V$ is the relative velosity). One does not want to get close
to it, but it still gives a scale for the small-distance scattering.
In the central 1~kpc region of our models the rms velocity of
particles is about $200~\kms$. With the mass of a particle of $2\times
10^5\Msun$, we get $b_{\rm min} = 2\times 10^{-2}$~pc. This is 2000
times smaller than our resolution of 50pc! And this still
overestimates the effect because we have many particles inside the
resolution element. At larger distance, say 5~kpc, the rms velocity in
the disk is smaller. Again, taking a typical rms velocity from our
models, we get $b_{\rm min} = 0.1$~pc. With the cell size of 88~pc
(resolution 160pc) we are still 2000 times away from $b_{\rm min}$. To
summarize, effects of close collisions are definitely small in our
models.

During the integration, spatial refinement is accompanied by temporal
refinement.  Namely, particles moving on each level of refinement are
advanced using the leap-frog scheme. When a particle moves to another
level, its velocity is re-interpolated. The time step decreases by
factor two with each refinement level.  This variable time stepping is very
important for accuracy of the results.  As the force resolution
increases, more steps are needed to integrate the trajectories
accurately.

\subsection{Code tests}

ART code has been extensively tested during the last $\sim 10$~years. 
A list of tests and details are presented in \citet{vk03}. Here we give a brief
summary of those which are important for our problems. Among other
tests, \citet{KKK97} tested the code using the spherical accretion
model. A small initial seed is placed in a homogeneous expanding
background. Shells surrounding the seed start to accrete on the seed,
which results in the built up of a very centrally concentrated
object. An analytical solution for this complicated process is known (the
Bertschinger solution). In the test, the density increases by almost 5
orders of magnitude. At the end, no detectable deviations of the
numerical solution provided by ART and the analytical solution were
found. This is an important test because it shows that the code
accurately treats collapsing systems. In the simulations of barred
galaxies, this collapse happens during the initial stage of formation
of bars (first $\sim$ 1 Gyr of evolution), when the density in the
central region increases few times.  Asymmetric collapse was tested
\citep{KKK97} using the Zeldovich approximation (plane wave
collapse). 

\citet{vk03}  present additional tests, which focus on a long-term stability of
equilibrium systems. These tests are relevant for later stages of
evolution of barred models, when the system changes its structure very
gradually. \citet{vk03} used 200,000 particles to set equilibrium for
a high concentration NFW halo. The test is actually very difficult for
the code. Orbits of particles in the NFW halo are very
elongated. Thus, particles move again and again from high density
regions with high force resolution to outer regions with low
resolution and back. Any numerical defects at boundaries of regions of
different resolution would result in evolution of the system or in
excessive scattering of trajectories.  No defects were detected for
gigayears of evolution. Using trajectories of individual particles,
\citet{vk03} estimated effects of the two-body scattering in the
simulations. The two-body scattering is clearly present. It is
consistent with predictions of the Chandrasekhar approximation (after
some scaling due to inhomogeneous distribution of density). Yet, the
time for the two-body scattering is very long: for a typical run with
$3\times 10^6$ particles the two-body relaxation time is $\approx
4\times 10^4$ Gyr. So, it is practically negligible.

Numerical two-body scattering is often discussed in conjunction with
effects of resonances in barred galaxies
\citep[e.g.,][]{weinberg2005}. Excessive scattering may prevent
accurate treatment of resonances, which are important factors in disk
galaxies. Formation of bars in dark matter, discussed below, is an
example of resonant interaction between stellar bars and the dark
matter.  \citet{daniel05} studied resonances in some of our models. A
large number of strong resonances were detected, including the familiar 
co-rotation, the inner and the outer Linblad resonances. Thus, the
two-body scattering in our simulations is small enough for the
resonance to be present and to be very important factors in the
evolution of the systems.

  After reaching a peak in their amplitude, bars in our models can
either grow, gradually decline, or remain in a steady state with
little change in the amplitude. For models with relatively low central
densities (e.g., \atwo\ and \khb) the evolution with time is robust
against changes of the numerical parameters: the timestep, the number
of the particles, and the force resolution. We rerun some of our
models doubling the number of particles while keeping the same {\it
small} timestep and the results do not change significantly. We also
run the \khb\ model with the GADGET code \citep{volker01}. In this case we used only
$10^5$ particles in the disk and $10^6$ total. The force resolution
was limited to 100~pc (Plummer softening). At the beginning of the simulation
the time-step was $\sim 1.5\times 10^5$ yr while by the end
of the evolution it had reduced to $\sim 4.4\times 10^4$ yr.
Figure~2 shows the evolution of the pattern speed $\Omega_p$ and the
bar amplitude $A_2$ for the model \khb\ run with the GADGET and the
ART codes.  Overall the agreement with the ART code is reasonably
good.  After 5~Gyrs of evolution the bar amplitudes and the bar
lengths are within 10 percent. The pattern speeds deviate not more
than 20 percent. Note that both simulations of the \khb\ model show
the same qualitative behavior: the pattern speed is barely changing
over 5~Gyrs of evolution and the amplitude of the bar is gradually
decreasing with time.

\section{The bar pattern speed}

Figure~2 shows the evolution of the bar pattern speed \omp\
and the amplitude of the bar $A_2$ for six models. 
To estimate \omp\ we first determine the orientation of the bar
computed applying iteratively the method of the tensor of inertia in the plane 
of the disk (see section 5 below). 
\omp\ is obtained subsequently by numerical 
differentiation: $\omp  = d\phi/dt$, where $\phi$ is the position 
angle of the bar. In practice, we use about
ten consecutive snapshots for which the increasing function $\phi$ 
is available and make a least-squares fit. Then \omp\ is given 
by the slope of the straight line. The bar amplitude $A_2$ is computed
as in \cite{vk03}. The curves are smoothed out in time by using
a top-hat kernel.
Models show what seems to be a generic feature
of bar simulations: an increase in $A_2$ is accompanied with
a decline in \omp\ and vice versa. A steady amplitude, on
the other hand, produces a roughly constant \omp. 
\begin{figure}[htb!]
\epsscale{1.0}
\plotone{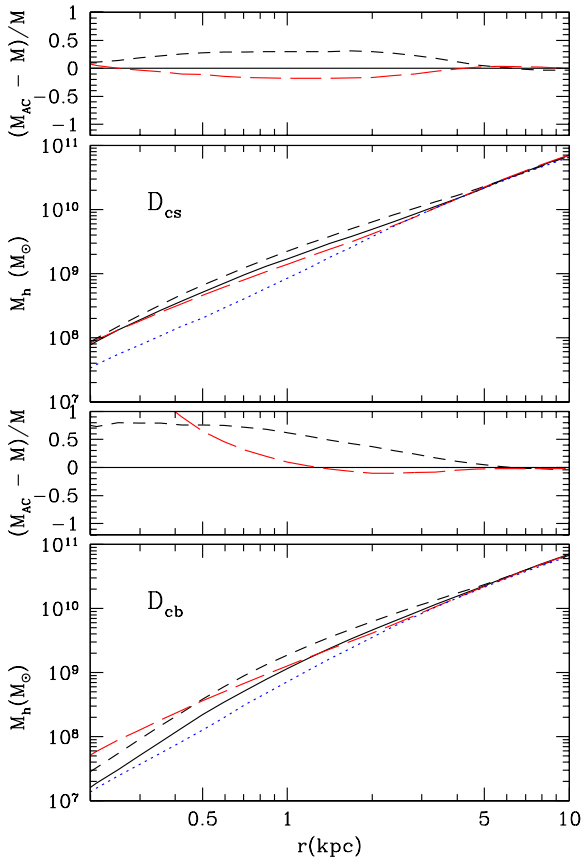}
\caption{The cumulative halo mass profiles for models \dcs\ and
\dcb. Initial profiles are shown by dotted curves.  The profile at
the last simulated epoch $t \sim 6\ \gyr$ is presented by full
curves.  Results of the standard and the modified adiabatic
approximations are presented by the short-dashed and long-dashed
curves. In the small panels we also show the relative mass differences
between the adiabatic approximations and the mass profiles as measured
in simulations.}
\end{figure}

\section{Adiabatic contraction}

The structure and the dynamics of the halo is altered by
the presence and evolution of the disk and vice versa. For example,
halo's inner shape changes 
from the assumed initial spherical configuration to a triaxial one. We will
see that in models, in which a strong bar developes, the halo becomes prolate.
 The $m =2$ structure of the halo,
the DM bar, which arises because of the bar instability of the disk, 
couples with the stellar bar. It is an important contributer to the 
stellar bar braking (see section 6 below). In this section, we study the changes
in its mass structure.
\begin{figure}[htb!]
\epsscale{1.0}
\plotone{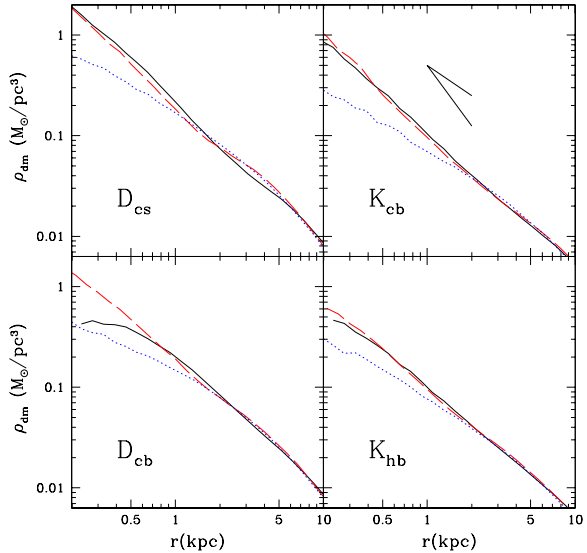}
\caption{Spherically averaged DM density profiles for models 
\dcb, \khb, \dcs\ and \kcb. Curves are coded as in
previous figure: initial profile 
(dotted curve), the profile at the end of the evolution, $t \sim 6\ \gyr$,
(solid curve), and the adiabatic approximation of Gnedin et al. (long-dashed curve). 
Noise in initial and final density profiles was reduced by averaging 
five consecutive timesteps. For comparison  in the top-right panel we 
show  power laws with the slope $-1$ and $-2$.}
\end{figure}

It is known \citep[e.g.,][]{am02,vk03,avila05} that as the stellar bar 
grows, the inner disk density increases. Is this density
growth accompanied with a corresponding central halo mass increase? 
If so, can it be modeled by the standard adiabatic contraction approximation?
This latter model assumes spherical symmetry, homologous contraction, particles 
in circular orbits, and conservation of the angular momentum \citep{blumenthal86,
rg87,gkkn04}. Under these assumptions, the final halo 
cumulative mass profile can be computed given the initial DM and barionic mass
profiles, $M_{DM}(r)$, $M_b(r)$, and the final barionic profile, $M_b(r_f)$:
\begin{equation}
\left[ M_{DM} (r) + M_b(r) \right] r =
\left[ M_{DM} (r) + M_b(r_f) \right] r_f.
\end{equation}

Given $r_f$ and, thus, $M_b(r_f)$, we solve the above equation for $r$ iteratively. 
The initial mass profiles as well as the final baryonic profiles are
taken from simulations. To reduce noise, both final and initial profiles are
smoothed out over few close timesteps.   \citet{gkkn04} 
tested recently the model in cosmological simulations and find that it 
systematically overestimates the central density. Notice that
since the orbits of DM particles are on average eccentric the assumption of 
circular orbits of the standard formalism breaks.  \citet{gkkn04} show  
that the model can be improved
if the adiabatic invariant is changed from $M(r)r$ to $M(\bar r)r$, where
$r$ and $\bar r$ are the current and orbit-averaged particle positions,
respectively. This modification approximately accounts for the eccentricity
of particle orbits.
It was found that $\bar r$ can be described by a power law:
\begin{equation}
\bar x = A x^w, \qquad x \equiv r/r_{vir},
\end{equation}
with small variations in the $A$ and $w$ parameters. We will test the new 
model adopting  mean values given by \citet{gkkn04}: $A = 0.85$ and $w = 0.8$.
\begin{figure}[htb!]
\plotone{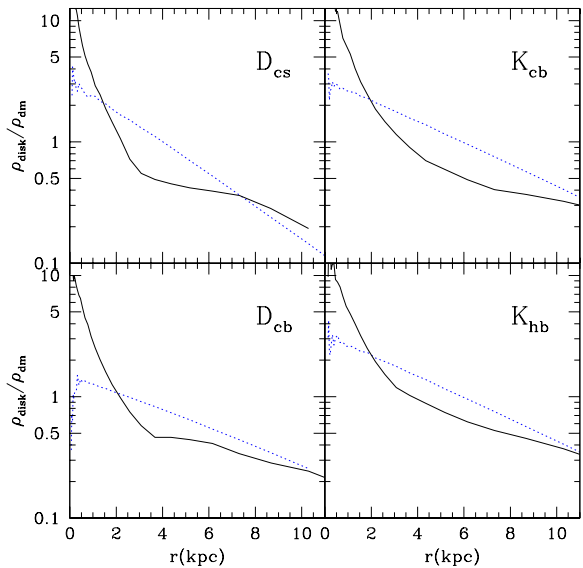}
\caption{The disk to dark matter ratio for models
 \dcb, \khb,  \dcs, and \kcb. The dotted curves represent the initial
models. The final models are shown with solid curves. The  increase in the disk
central density  is produced at the expense of the density
in middle region. This effect is slightly more pronounced in 
in the cold- and disk-dominated models \dcs, and \kcb. } 
\end{figure}

Figure~3 shows the initial and the final halo mass profiles for models
\dcs\ (top panel) and \dcb\ (bottom panel). The mass profile in the
model \dcb\ barely changes, while in the model \dcs\ the mass at
0.5~kpc, for example, has increased substantialy -- by a factor of
2.5. Overall, the central mass increase seem to be reproduced better
by the modification introduced by \citet{gkkn04}. The standard formula
overestimates the central density, which agrees with the results of Gnedin et
al. for cosmlogical halos.

Figure~4 shows the dark matter density profiles of models \dcs\ and \kcb\
(cold and disk-dominated models), and models \khb\ and \dcb\ (hot and
dark-dominated models). The adiabatic density profiles are
computed by differentiating the corresponding mass profiles. We
decided to show only the predictions by the adiabatic contraction
formalism developed by \citet{gkkn04} because the standard 
adiabatic contraction formula does not produce better results.
In order to reduce the shot noise,  we average five
consecutive snapshots in both final and initial profiles. As was already 
noticed in the cumulative mass profiles (Fig. 3), the increase in the inner
halo density is higher in disk-dominated and cold models (\dcs\ versus
\dcb\ or \kcb\ versus \khb). The inner slope is steeper than
$\gamma=-1$ in all models except in the model \dcb. The rather flat inner density
slope  of the model \dcb\ (see left-bottom panel) is related with the fact
that  the inner density profile of the stellar disk is also flat
in this model.

To summarize, we find that the increase in the halo central density
is compatible with the adiabatic compression. In particular, we
find that the inner halo density increase is better reproduced by
the formalism developed by \cite{gkkn04}. In models
in which this density increase is smaller (model \dcb, for instance), 
the adiabatic compression tends to overestimate the DM density 
in the central region.

Figure~5 gives additional information on the changes in the dark
matter.  It shows the ratio of the disk to the halo densities as a
function of radius.  At the end of the evolution in all models the
stellar component strongly dominates the central $\sim 1$~kpc region.
\begin{figure}[htb!]
\epsscale{1.0}
\plotone{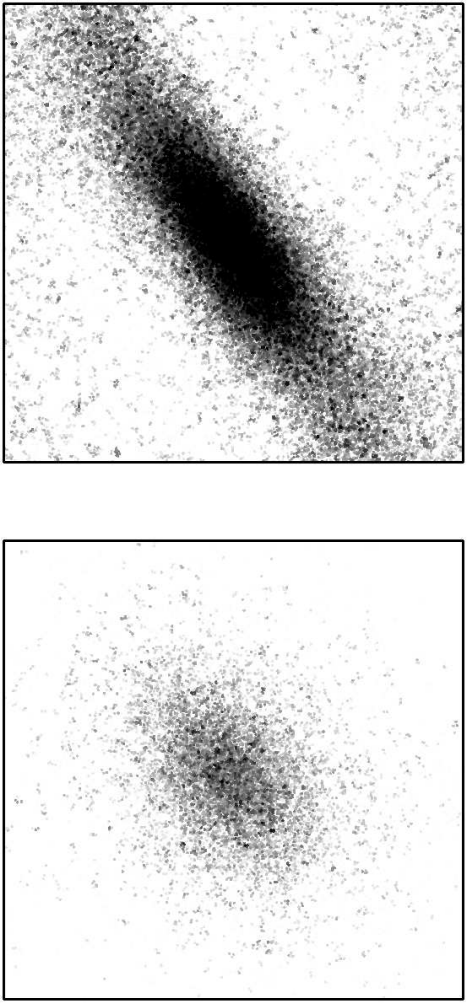}
\caption{Distribution of disk (top panel) and dark matter particles
(bottom panel) inside a box of 10~kpc across
for model \dcb\ at $t \sim 6 \gyr$. The figure shows the
face-on projection of the particles. The halo bar is clearly seen in
the bottom panel.  To improve the contrast, we have color-coded
particles on a gray scale according to their local 3D density (a
pgplot code kindly provided by A. Kravtsov) and plotted only DM
particles with the $z$- component of velocity $|v_z| < 100\ \kms$.}
\end{figure}

\section{Shape of the dark matter halo}

The ellipticity of different components (DM or stellar) as well as the
directions of the principal axes are determined by iteratively
diagonalizing the inertia tensor.  We start by finding the tensor of
inertia for all particles inside a spherical shell of given radius
$R$. We can take either stellar or DM particles depending on what
component we study. We then find principal semi-axes $a > b > c$,
and angles of the tensor. In the next iteration, we find the modified
inertia tensor $I_{ij}$ for particles inside an elliptical shell with the
orientation given by the inertia tensor on the previous iteration and
with the semi-axes $R$, $(b/a)R$, and $(c/a)R$:
\begin{eqnarray}
I_{ij} &=& \sum_k \frac{x_{i,k} x_{j,k}}{d_k^2}, \\
 d_k^2 &\equiv& x_{1,k}^2 + \frac{x_{2,k}^2}{q^2}
+ \frac{x_{3,k}^2}{s^2}, \\
q &\equiv& b/a, \qquad s \equiv c/a. 
\end{eqnarray}
\noindent Here the sum is taken over all particles inside the elliptical shell.
In equations (8-10) $x_{i,k}$ ($i=1,$ 2, and 3)
are the coordinates of the particle $k$ with respect to the center of
the disk-halo system and $d_k$ is the elliptical distance; $s$ is
the short-to-long axis ratio, and $q$ is the intermediate-to-long
axis ratio. The center of the system is found iteratively as the
center of mass of a sphere containing maximum mass. The radius of the
sphere is equal to the disk scale length $R_d$. The elliptical bins 
are chosen so as they have roughly the same number of particles and
are less than $1~\kpc$ width. As expected, the minor axis lies almost perpendicular to
the plane of the disk.

Figure 6 highlights the presence of the stellar and the DM bars in the
model \dcb, which has  a strong bar. The stellar bar shown in the top panel has a
rectangular-like shape, while the halo bar shown in bottom panel is
more round. Notice that both bars have their major axes pointing to
approximately the same direction.
\begin{figure}[htb!]
\epsscale{1.0}
\plotone{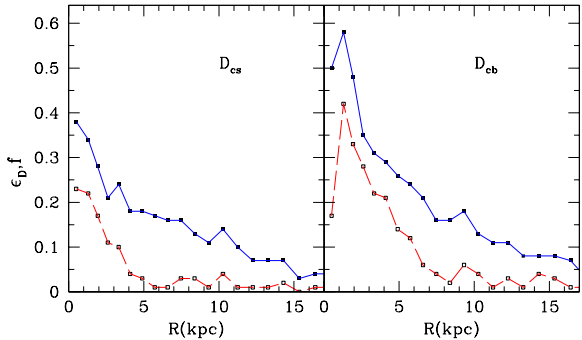}
\caption{Flattening $f \equiv 1 - c/a$ (solid curve) and ellipticity
in the plane of the disk $\epsilon_D \equiv 1 - b/a$ (dashed curve) of
the DM distribution, where $a,b,$ and $c$ are the major, intermadiate,
and the minor axes.  We show models \dcs\ (left panel) and \dcb\
(right panel) at the last measured epoch $t \sim 6\ \gyr$. 
 The differences between the weak bar
model \dcs\ and the strong bar model \dcb\ are clearly seen.}
\end{figure}

Figure~7 shows the flattening, $f \equiv 1 - s =1-c/a$, and the
ellipticity of the dark matter in the plane defined by major and
intermediate axes, $\epsilon_D \equiv 1 -q=1-b/a$, as a function of
distance $R$ to the center for models \dcs\ and \dcb\ at the last
simulated epoch $\sim 6\ \gyr$ (hereafter $f$ and \eD\ are referred to
as ellipticities). The plot shows that DM bars are typically triaxial.
In models with a strong stellar bar, as in the model \dcb, the DM bar
can be more prolate than oblate. As shown in Figure~7, axial ratios
increase with increasing radius (thus, the system is getting more
round). For example, at 15~kpc models \dcs\ and \dcb\ show comparable
small ellipticities, while in the central regions the models have very
different flattening.

Evolution of the DM ellipticities with time is shown in Figure~8.  By
comparing Figure~2 with Figure~8 we find similarities of behavior of
the stellar bar amplitude with the evolution of DM bar
ellipticity. Indeed, in models \dhs\ and \khb\ the maximum in the
stellar bar amplitude $A_2$ and the maximum of ellipticity
$\epsilon_D$ are reached approximately at the same time.  Moreover, in
the model \dcb\ the second period of the stellar bar growth is mirrowed
by an increase in the ellipticity of the DM bar.  The gradual decline of
$A_2$ in models \dhs, \kcb, and to smaller degree in the model \khb\,
is imitated by a decline in \eD.

 At the onset of the evolution, the sole presence of the disk produces
a flattening of the DM halo in the central region. The subsequent
evolution of the shape of the DM halo depends on the model. For example, in
model \dhs\ the ellipticities first increase as the stellar bar growths.
Then they decline maintaining a figure of constant triaxiality. On the
other hand, in the model \dcb\ the flattening $f$ stays approximately constant
after the initial period of stellar bar growth. This latter effect,
along with the growth of ellipticity \eD, produces in  models \dcb\ and
\atwo\ (not shown in the figure) a near prolate central DM halo by the
end of the evolution.  In summary, Figures~2, 6, and 8 show a clear
indication of the disk-halo coupling (for details see section 6).
\begin{figure}[htb!]
\vspace{7.8cm}
\includegraphics{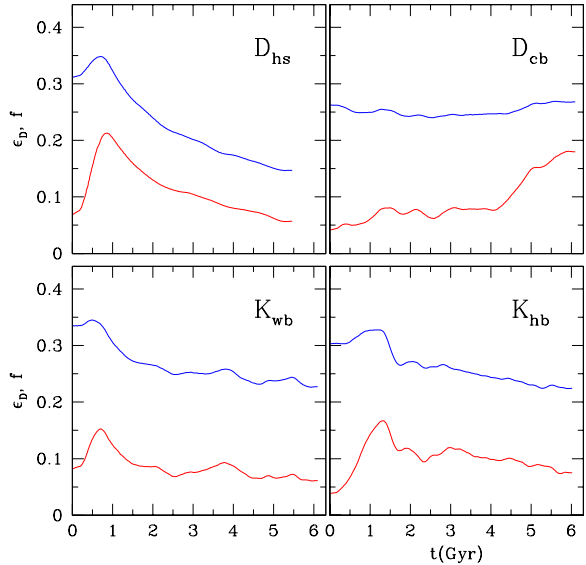}
\caption{Evolution of the DM ellipticities, $f$
(top) and \eD\ (bottom) for four of our seven models. Ellipticities
are evaluated at $r = R_d$, where $R_d$ is the scale length of the
corresponding model.}
\end{figure}

\subsection{Bar orientation}

We use the position angle, PA, of the stellar bar and the orientation
of the DM bar measured at $r = R_d$ to build the evolution of the bars
phase difference, $\Delta \phi$. Figure~9 shows the evolution of
$\Delta \phi$ for four of our seven models.  At $R_d$ the bar
alignment within $\sim 10^\circ$ is reached soon after the initial
stellar bar growth (the DM bar trails the stellar bar). The start of
the second period of stellar bar growth in model \dcb\ coincides with
a small increase in $\Delta\phi$ -- the DM bar gets a larger lag. This
renewed disalignment quickly dies out and the bars again rotate in
phase.  More significantly, this second period coincides with the
period of a strong influence of the stellar bar on the halo, which is
reflected in the more extended bar aligment (see top-left panel,
dot-dashed curve). These two periods of stellar bar growth and bars
alignment also coincide with the two periods in which a higher angular
momentum lost-rate is observed (see Figures~11 and 12).

The phase difference $\Delta \phi$ of the bars is related to the torque
experienced by the stellar bar:  the torque slows down the bar pattern
speed. In the model \atwo\ (see Figure~2) the oscillations seen in \omp,
and to some degree also in $A_2$, are very likely coming from the
strong oscillations seen in $\Delta\phi$ in this model. The onset of
the late period of bar growth coincides with a decline of about $40^\circ$
in $\Delta\phi$: the $m=2$ halo structure is quickly re-oriented 
toward the direction of the stellar bar. Moreover, the rapid growth of the stellar bar
in the model \dcs\ (see Figure~2) produces a DM bar that very soon
aligns with the stellar bar.

\begin{figure}[htb!]
\vspace{7.8cm}
\includegraphics{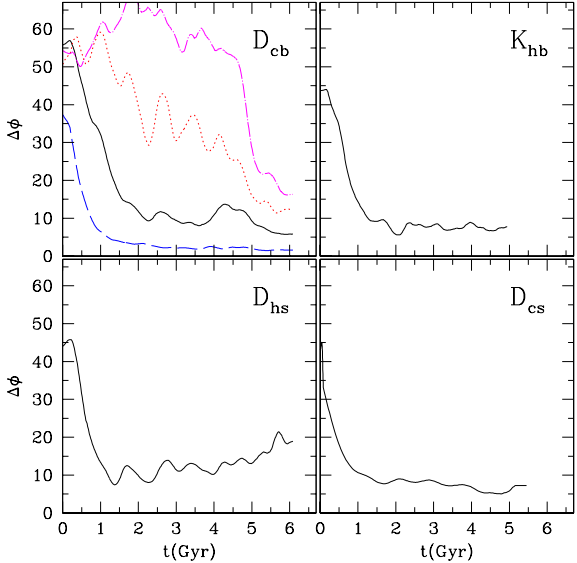}
\caption{Evolution of the phase difference between the stellar and the
DM bars for four  models at radius $R_d$ (solid
curves). For the strong bar model \dcb\ we also show the evolution of
$\Delta \phi$ at other radii: $R= 0.5R_d$ (dashed curve), $1.5R_d$ 
(dotted line), and $2.0R_d$ (dot-dashed curve). The curves show that
the  phase difference is an increasing function of 
radius. Interestingly, the two periods in which $\Delta \phi$ drops
(see solid and dot-dashed curves) seem to coincide with the two
periods in which a higher angular momentum lost-rate is observed.
At $R = R_d$ an alignment within $10^\circ$ is observed
in models that end up with strong and moderate 
stellar bars by the end of evolution.}
\end{figure}

\section{Angular momentum}

\begin{figure}[htb!]
\vspace{11.0cm}
\includegraphics{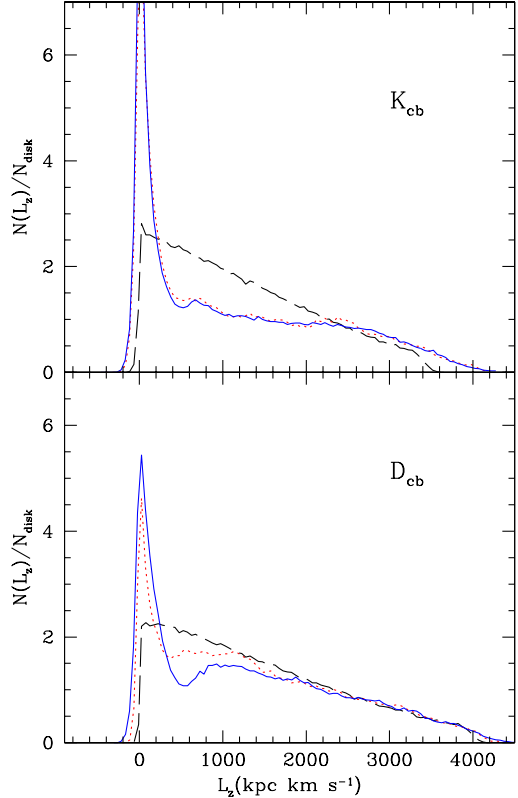}
\caption{Distribution of the $z$-component of the angular momentum
of the stellar particles for models \kcb\ (top panel) and \dcb\
(bottom panel), shown at different epochs: initial (long-dashed curve),
at $t \sim 1.6\ \gyr$ (dotted curve), and at $t \sim 5\ \gyr$, 
(solid curve). Particles with low (high) angular
momentum are preferentially at small (large) radii. The distributions
are qualitatively similar to those shown by \cite{vk03} (see their
Figure 12): a peak at low $L_z$, which corresponds to the
bar, a valley at intermediate values of $L_z$, and an excess of 
of particles with very large angular momentum. Notice that in
the weak bar model \kcb\ the distribution barely changes for the period $1.6-5.0\ \gyr$, 
while the redistribution of angular momentum continues in the strong bar model \dcb.}
\end{figure}

The bar growth is intimately related to its slowing down. The rate 
of change of \omp\ correlates with the initial conditions. For example, 
the $A_2$ maximum is reached earlier in  disk-dominated models 
(compare \dcs\ with \dcb), or in cold disk models (\kcb\ versus \khb). During
the bar braking the angular momentum is transfered from the disk to the halo.

Figure~10 shows the distribution of the disk $L_z$ component of the
angular momentum for two models, the strong bar model \dcb\ and the cold model
\kcb, one of the models with a strong decline in bar amplitude. 
We show the distribution of $L_z$ at three moments of time: initial, at $\sim 1.6~\gyr$,
and at $\sim 5~\gyr$. Disk particles in the intermediate-$L_z$ zone
lose angular momentum and move to the low-$L_z$ region close to the
center. At the same time, the number of particles with very
high-$L_z$, which are preferentially at larger radii, increases. Note
that the number of intermediate-$L_z$ particles in the model \kcb\ barely
changes from $t \sim 1.6$ to $\sim 5\ \gyr$, while this is not the case
in the model \dcb.  This behavior is compatible with the much higher
angular momentum loss  by the model \dcb\ during
this period of time shown in Figure 11.

The evolution of the relative change in $L_z$
is plotted in Figure~11, where we show four models: three cold models
\dcs, \kcb, and \dcb\  and one hot and disk-dominated model \khb. 
In the \kcb\ model, the amount of angular momentum
lost by the disk is very small during the declining phase of the bar 
amplitude. Moreover, the disk in models \dcs\ and \dcb\ has lost, 
by the end of the evolution, about 10\% of its angular momentum.

\begin{figure}[htb!]
\vspace{7.8cm}
\includegraphics{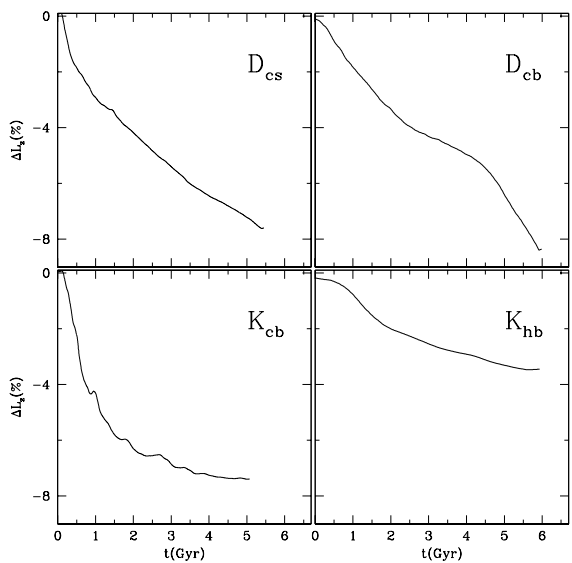}
\caption{Evolution of the disk angular momentum for four
models. We show two models, \dcb\ and \dcs, that end up with 
strong bars at the end of the simulated period and two other models, \kcb\ 
and \khb. Model \dcb\ presents an inflexion point or zone, in which
the rate of angular loss changes its decreasing trend to an increasing
one. This time coincides with the second period of bar growth.}
\end{figure}

It is interesting to estimate what fraction of the torque experienced
by the stellar disk is coming from the interaction between the dark
matter bar and the stellar bar. Unfortunately, the torque is difficult
to estimate accurately because of the complex structure of the bars: the bars
change shape, mass, and length as the evolution proceeds.  Still, we
can start with simple scaling relations. The torque should be
proportional to masses and radii of the bars. It is also proportional
to the strength of the bars, which can be measured by ellipticities
$q=1-b/a$, where $a$ and $b$ are the major and minor semi-axis in the
plane of the disk.  The angle $\Delta\phi$ between the bars also
affects the torque. From the geometry of the problem, we expect
that the torque is zero, when the angle is either zero or $90^\circ$.

In order to test the scaling relations, we estimate the torque between
two elliptical-shape objects, which roughly mimic our bars. Boundary
of each object is assumed to be a triaxial ellipsoid with semi-axes
$a > b >c$: $(x/a)^2+(y/b)^2+(z/c)^2=1$.  For the ellipsoid representing
the stellar bar we use $c/a = 0.1$. For the dark matter bar
$c/a = 0.5$. The other axial ratio $b/a$ was varied for both ellipsoids to
test its effect on the torque. We also varied the angle between the
bars.  The density was chosen to fall exponentially with the distance
mimicking the decline of the density in bars measured in our N-body models.
The ellipsoid representing the dark matter bar was twice shorter as
compared with the stellar bar. We use Monte Carlo
method to estimate the torque between the ellipsoids. We find that the
torque $\tau$ between the ellipsoids scales as $\tau \propto
\epsilon_{\rm DM}\epsilon_{\rm stellar}\sin(2\Delta\phi)$, which is
what one naively expects. Once the parameters of the ellipsoids are
fixed, the model also provides absolute value of the torque. We used
parameters of bars in the model \dcb\ at 3~Gyr to roughly estimate
the torque. Indeed, we find that the expected torque between the bars
and the measured torque in the model \dcb\ are quite close (within
30\%). This is reassuring, but unfortunately, it is difficult to press
this issue because matching the masses and the density distributions
with bars in real simulations is difficult. Instead, we decided to use
only the scaling relation for the torque. In this case we use the
amplitude of the stellar bar $A_2$ as a measure of the bar
strength. The amplitude of the expected torque is fixed to have the
observed value at one moment in our models: at 3~Gyr.
 
We tested three of our models.  In Figure~12 we show the rate of
angular momentum loss of the disk as a function of time (top panels)
for models \dcb\ and \khb.  In the bottom panels
we show the evolution of the quantity $\tau = A_2 \epsilon_D
\sin(2\Delta \phi)$ measured at $r = R_d$ and normalized at 3~Gyr.
$A_2$, $\epsilon_D$, and $\Delta \phi$ are the amplitude of the disk
bar, the ellipticity of the DM bar, and the phase between the bars,
respectively, introduced in previous sections. With the exception of
the first $\sim 1$~Gyr (model \dcb), the observed torque $dL_z/dt$ and the
estimated torque between the bars are remarkably similar. After initial
stage, at $t\approx 1-2.5$~Gyr the torque declines. The second growth 
of the stellar bar at about 4~Gyr in model \dcb\ produces an increase 
in $dL_z/dt$ at later epochs.

We also find an agreement between the measured and estimated torques
for other models. We note that none of the quantities in the
estimate of $\tau$ can {\it separately} explain even qualitatively the
time behavior of $dL_z/dt$. Indeed, the decline in the torque from
1.5~Gyr to 3~Gyr is mostly due to decreasing of the angle between the
bars. Yet, the steep increase in $dL_z/dt$ at 4~Gyr in model \dcb\
is not related
with the angle: the angle keeps decreasing. The increase is mostly due
to the growth of the amplitude of the stellar bar, which is
accompanied by an increase in the ellipticity of the dark mater bar
(see Figure~8). In spite of general agreement of the torques, there
are some deviations.  At early stages ($t<1$~Gyr) the scaled value of
$\tau$ is too small as compared with the measured torque
in the cold model \dcb. This is
hardly surprising because the disk in this model develops instabilities very
quickly. In addition, it was difficult to set the very central $\sim
200$~pc of the disk initially in equilibrium. So, irregular
non-axisymmetric waves are present almost from the very
beginning. These waves result in $\sim 1-2\%$ loss of the disk angular
momentum. At around 0.1~Gyr the disk develops strong spiral arms,
which likely account for a large fraction of the torque at that
time. As the spiral arms die and the bar grows, the torque starts to
be dominated by the interaction between the dark matter and the
stellar bars, and the $\tau$ approximation works better.

Judging by the similarities between $dL_z/dt(t)$ and
$\tau(t)$, we find it plausible that the torque between the stellar
and the dark matter bars is responsible for a large fraction of
angular momentum transfer between the disk and the dark matter.

\begin{figure}[htb!]
\vspace{7.8cm}
\includegraphics{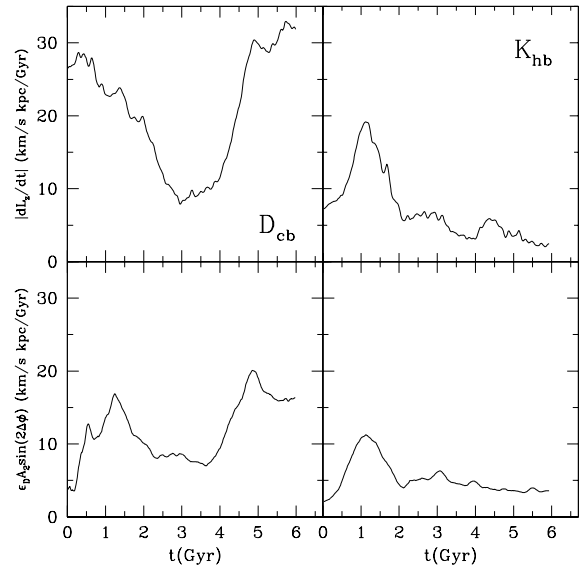}
\caption{The top panels show the specific torque $dL_z/dt$ experienced by
the stellar component in models \dcb\ and \khb\ with a strong 
and a moderate stellar bar, respectively. The
bottom panels show the evolution of an estimate of the torque due only to
interaction  between the stellar and the dark matter bars: $\tau = \epsilon_D
A_2 \sin(2\Delta \phi)$, where $\epsilon_D$, $A_2$, and $\Delta \phi$
are the ellipticity of the halo bar, the amplitude of the stellar bar,
and the lag angle, respectively.  Torque $dL_z/dt$ is reasonably well
reproduced by interaction between the  bars during most of evolution.}
\end{figure} 

\section{Discussion and conclusions}

Just as many other groups in the field \citep[\eg][]{lia05a,lia05b,ss04,ds98} we 
find that our disk models develops strong, moderate,
and  weak bars. 
They show what seems to be a generic 
feature of bar simulations: the growth in their amplitude is accompanied with 
a decrease in their pattern speed. The role that the halo plays in this phenomenon 
appears to be crucial. The dark matter ellipsoidal figure, 
the halo bar, that arises due to 
the presence of the disk and stellar bar, exerts a torque on the
stellar bar that slows down this latter. It is thus not 
surprising to find that the first period of bar growth 
(the only one in some models)
coincides with a decline in the phase difference between the disk and the
halo bar. Interestingly, an opposite trend seems to exist as well: bars
rotate faster as they weaken. This is clearly seen in models \kcb\ and \dhs. 

Disk particles located in the middle 2-8 kpc region transfer angular
momentum to the halo and to the outer disk region. In particular, this
effect would explain the increase of the central stellar density seen,
for example, in Figures~8-9 of \citet{vk03} \citep[see also][]{avila05}. 
We have seen in section~4 that this disk density increase is 
accompanied by a corresponding
central halo density increase.  We apply the adiabatic contraction
formalism to study whether the compression of the halo could be modeled
by it. We find that the modified formalism by \citet{gkkn04}
reproduces relatively well the inner density growth whereas the standard
formula typically overestimates the density increase.  

We compute the ellipticities of the halo bar in a plane parallel, \eD, 
and perpendicular, $f$, to the disk plane. The ellipticities decrease and the halo 
bar becomes more spherical as the radius increases. In some models, after 
an initial increase the flatness stays roughly constant or decreases during the
disk evolution. On the other hand, in models with strong bars like model \dcb,
the ellipticity \eD\ has a second period of increase, which agrees with the second growth period
of the stellar bar. As a result, at end of the evolution the inner dark matter halo
ends up  with a near prolate-like shape.  Our results roughly agree with
those mentioned in \citet{lia05a,lia05b}, these later being preliminary.

In all models, the inner halo ($r \simless 5 \kpc$) is flattened, with
the minor axis pointing perpendicular to the disk plane. The alignment
of the minor axis with the disk axis is compatible with the results by
\citet{bailin05}. In cosmological simulations of disk galaxy formation
\citet{bailin05} find that the the inner halo ($r \simless 0.1 \rv$)
orients its minor axis parallel to the disk axis, regardless of the
orientation of the outer halo. As they explain, the misalignment
between the two halo regions should be taken into account when
modeling tidal streams in the halos of disk galaxies.

\citet{loa92} found that the presence of a stellar bar in the center of our 
Galaxy does not strongly affect the destruction rates of globular clusters. The
results of \citet{loa92} or of similar studies \citep[\eg][]{pmm04} might 
change in light of our findings: halo bars could enhance the dynamical 
effect of bars on globular clusters distribution.

We use the same method of the tensor of inertia to study the evolution
with time of the orientation of the halo bar relative to the disk
bar. In all models, an alignment within $\sim 10^\circ$ is obtained
after the initial bar growth. In some models, for example model \khb,
the bars stay aligned during the whole evolution. It is interesting to
see that in the model \dcb\ the periods of alignment roughly agree
with the two periods of stellar bar growth. In turn, the periods of
bar growth are also the periods of higher angular momentum lost rate.

As  \citet{vk03}, we also find that the angular momentum transfer from the disk to 
the halo is rather modest: it is less than 10\%. As it was shown by
Valenzuela \& Klypin, this is partly due to the fact that a significant fraction of the 
disk angular momentum is not lost but redistributed: outer regions absorb 
part of the angular momentum of intermediate regions. 

\acknowledgments

P.C. acknowledges support by CONACyT grant 36584-E. 
A.K. acknowledges support by NSF and NASA grants to NMSU.
O.V. acknowledges support by the NSF ITR grant NSF-0205413.
We thank the anonymous referee whose helpful comments and
suggestions improved some aspects of this paper.

\end{document}